\documentclass[pre,aps,twocolumn,showpacs,floatfix,10pt,superscriptaddress,longbibliography]{revtex4-1}
\usepackage[dvips]{graphicx}
\usepackage{subfigure}
\usepackage{parcolumns}
\usepackage{amsmath}
\usepackage{amssymb}
\usepackage{color}
\usepackage{hyperref}

\newcommand{\MeijerG}[7]{G \begin{smallmatrix} #1 & #2 \\ #3 & #4 \end{smallmatrix} \left(#7 \middle\vert  \begin{smallmatrix} #5 \\ #6 \end{smallmatrix} \right) }

\begin{document}

\title{Inhomogeneous parametric scaling and variable-order fractional diffusion equations}

\begin{abstract}
We discuss the derivation and the solutions of integro-differential equations (variable-order time-fractional diffusion equations) following as continuous limits for lattice continuous time random walk schemes with power-law waiting-time probability density functions, whose parameters are position-dependent. We concentrate on subdiffusive cases and discuss two situations as examples: A system consisting of two parts with different exponents of subdiffusion, and a system in which the subdiffusion exponent changes linearly from one end of the interval to another one. In both cases we compare the numerical solutions of generalized master equations describing the process on the lattice with the corresponding solutions of the
continuous equations, which follow by exact solution of the corresponding equations in the Laplace domain with subsequent numerical inversion using the Gaver-Stehfest algorithm. 

\end{abstract}

\author{Philipp Roth}
\email{roth@physik.hu-berlin.de}
\affiliation{Institute of Physics, Humboldt University Berlin, Newtonstra{\ss}e 15, 12489 Berlin, Germany}
\author{Igor M. Sokolov}
\email{igor.sokolov@physik.hu-berlin.de}
\affiliation{Institute of Physics, Humboldt University Berlin, Newtonstra{\ss}e 15, 12489 Berlin, Germany}
\affiliation{IRIS Adlershof, Humboldt University Berlin, Zum Gro{\ss}en Windkanal 6,
12489 Berlin, Germany}

\maketitle

\section{Introduction}
\label{sec:Int}

The kinetic equations with partial fractional derivatives in temporal or spacial variables have recently attracted broad attention and 
are widely used for description of relaxation processes in complex media, see e.g. \cite{MetzKla,SoftMatt,Liang,klagessok} for reviews. 
If in addition to a property of being locally ``complex'', the system shows strong inhomogeneity on
larger scales, the parameters of the ensuing equations, including the order of the corresponding derivatives, may get position-dependent. 
A variant of such a situation starting from the continuous time random walk (CTRW) with position-dependent power-law waiting time distribution was
studied in \cite{frac}. The approach was followed in several subsequent works \cite{para,straka2018variable,fedotov2019asymptotic,fedotov2012subdiffusive}. A specific situation of 
a system consisting of two media with different properties of (sub)diffusion in contact with each other was under especially extensive 
consideration \cite{shkilev2013boundary,kosztolowicz2018model,kosztolowicz2008subdiffusion,kosztolowicz2015random,korabel2011boundary,marseguerra2006normal}. In this special case the emphasis lays on derivation of the matching conditions at the 
boundary of the two media. 

In the present work we discuss in some detail the derivation of the variable-order fractional diffusion equations for inhomogeneous 
media from the corresponding generalized master equations for CTRWs and the corresponding scaling limits, and show how these equations naturally appear under the limiting transitions. Two specific situations are considered in some detail:
continuous changes in the parameters of the local waiting time distributions, as discussed in \cite{fedotov2019asymptotic}, and the abrupt change at the border of the medium, 
in which case the corresponding matching conditions emerge naturally from the requirement of the existence of the scaling limits, as discussed in \cite{shkilev2013boundary}. 
We moreover propose a numerical procedure to solve the corresponding equations, and check our analytical results against numerical
solutions. We also present the (semi-)analytical and approximated solutions of the corresponding equations for the two examples.

\section{From the master equation to generalized diffusion equation}

Since the generalizations to higher dimensions are evident, we focus on a one-dimensional situation in the present work.
We concentrate on the case of CTRWs on a regular lattice with lattice spacing $a$. The off-lattice situations can be discussed by similar methods but require a more involved analysis.
We start from the generalized master equation (GME) for CTRW on a regular one-dimensional lattice, see Eq.(5.15) of Ref. \cite{steps},
\begin{equation}
\label{gme}
 \frac{d}{dt}p_i = \frac{d}{dt}\int_0^t dt' M_i(t-t') \left[\frac{1}{2} p_{i-1}(t') +  \frac{1}{2} p_{i+1}(t') - p_i(t')\right] 
\end{equation}
with $i$ numbering the sites and $M_i$ being the kernels of site-dependent integro-differential memory operators $\hat{M_i}$. This equation, with two neighbors for each site, is
valid for the internal sites of the lattice. For boundary sites of a finite lattice, only one neighbor, to the right or to the left
of the corresponding site, is present. 

The kernel of the memory operator takes the simplest form in the Laplace domain \cite{steps,frac}: 
\begin{equation}
 M_i(s) = \frac{\psi_{i}(s)}{1-\psi_{i}(s)},
\end{equation}
so that for the internal sites of the system
\begin{eqnarray}
s p_i(s) -&& p_i(t=0) = \frac{1}{2} \frac{s \psi_{i-1}(s)}{1-\psi_{i-1}(s)}p_{i-1}(s) \label{eq:Main} \\
&& + \frac{1}{2} \frac{s \psi_{i+1}(s)}{1-\psi_{i+1}(s)}p_{i+1}(s) -\frac{s \psi_i(s)}{1-\psi_i(s)} p_i(s). \nonumber
\end{eqnarray}
The function $\psi_i(s)$ is the Laplace transform of the waiting time density $\psi_i(t)$ at a site $i$. For the case of the usual master equation, the function $\psi_i(t)$
is exponential, $\psi(t) = \tau_i^{-1} \exp(-t/\tau)$; for typical cases of anomalous diffusion it is a power law, $\psi_i(t) \sim \tau_i^{-1} (t/\tau_i)^{-1-\alpha}$.
In both cases $\tau_i$ represents the characteristic waiting time at a site $i$ (in the first case it is simply the mean waiting time, 
in the second case it is of the order of the median waiting time). 

In what follows we discuss the three following situations: (a) the homogeneous system (all $\psi_i$ are the same), which serves as a starting point of our
considerations, and the two specific situations pertinent to heterogeneous systems: (b) the situation with the boundary (there are two kinds of $\psi_i$ for sites to the left and to the right of the boundary),
and (c) the situation, when $\psi_i(t)$ are gradually changing with position, both mentioned in Sec. \ref{sec:Int}. 
For the second situation, we discuss the question about the matching condition on the boundary and present a solution to the corresponding equation. For the third one we concentrate on the question about the form of the corresponding continuous equation,
and on its solution. The corresponding questions were posed and partially answered in the 
previous works \cite{shkilev2013boundary,kosztolowicz2018model,kosztolowicz2008subdiffusion,kosztolowicz2015random}; the present work unifies the approaches, shows how the continuous equations and the matching conditions arise naturally under the 
limiting transition to continuum, presents analytical solutions for some special cases, and proposes an effective numerical scheme.

\subsection{Homogeneous situation}
In the homogeneous case one can consider the usual scaling limit of large scales and long times. 
For the case of fractional diffusion in an infinite homogeneous medium this scaling procedure is discussed in detail in Ref. \cite{BarSok}.

In the simplest case corresponding to an infinite homogeneous system, one can represent the procedure as follows: One assumes that at longer times and at larger scales the solution of the Master equation $p_i(t)$ can be interpolated by a 
sufficiently smooth function $p(x,t)$ with $p(ai,t) = p_i(t)$. One rescales the coordinate $x \to c x$ and the time $t \to c^z t$,
considers $c \to \infty$, and looks for such a value of  $z$ that the solution takes a scaling form
\begin{equation}
 p(x,t) = \frac{1}{t^\beta} f \left(\frac{x}{t^\beta} \right)
\end{equation}
where the function $f(\xi)$ neither vanishes identically nor tends to a delta-function.  This corresponds to $\beta = 1/z = \alpha/2$.
The procedure represents the standard asymptotic scaling at long times, i.e. at times which are much larger than the intrinsic time scale $\tau$
of the system, and at length scales which are much larger than the intrinsic length scale $a$.

The situation discussed in the present work is different: the asymptotic scaling described above is useless in inhomogeneous situations 
for generalized master equations, where the final asymptotic solution is concentrated on a single site with the 
smallest value of $\alpha$ \cite{fedotov2012subdiffusive,fedotov2019asymptotic}, i.e. does tend to a delta-function.
We are not seeking for the final, but for the intermediate asymptotics of the PDF (provided it exists), and essentially even not for the form 
of the solution, but for the form of the equation defining it. Therefore, the term ``intermediate asymptotics'' is used above in a very similar 
but not in exactly the same sense as it is used in classical works \cite{BarZel,Barenblatt}, when the equations are given, and their solutions are sought for. 

The existence of an equation for intermediate asymptotics implies that at some intermediate times $t$
(i.e. in the interval $t_{\min} \ll t \ll t_{\max}$, with $t_{\min} \sim \tau$ and $t_{\max}$ depending on the particular situation at hand)
the behavior of $p_i(t)$ defined at lattice sites $i$ can be well approximated by a sufficiently smooth function $p(x,t)$ of the
coordinates of the sites given by a solution of some partial (integro-)differential equation, which does not explicitly contain the lattice spacing $a$.
In the domain, where this approximation is applicable (i.e. at intermediate times) this function should not change considerably on scales of the order of 
lattice spacing, and therefore has to be invariant under changes of this spacing provided the macroscopic parameters of the system
are kept constant (the precise meaning of this statement will be discussed in detail below). 
Therefore, the idea is, fixing the actual time $t$, to change the \textit{internal parameters} $a$ and $\tau$ of the system so that 
this fixed $t$ fulfills $t \gg \tau$ and that the ensuing solution (truthfully approximating $p_i(t)$ at the nodes of the initial lattice) 
does not change considerably on the scales of this new $a$ -- we will call this procedure parametric scaling, representing a kind
of ``van-Hove'' or ``fluid'' limit, see \cite{PFO}. This is e.g. a scaling limit making a simple random walk to be a standard Brownian motion (by virtue of the Donsker's functional central limit theorem),
when the transition $a \to 0$, $\tau \to 0$ is made when keeping the diffusion coefficient $D \propto a^2/\tau$ constant, i.e. $a \to \lambda a$, $\tau \to \lambda^2 \tau$.
In this case the master equation transforms into a Fokker-Planck equation. 
For CTRW in a homogeneous setting, the concept of parametric scaling was first applied in Ref. \cite{Gorenflo}.
Note that while the asymptotic scaling corresponds to a coarse-graining procedure, the parametric scaling is a kind of a "fine-graining" one.

Thus, we assume that our discrete-space result $p_i(t)$ can be well approximated by a function $p(x,t)$ which is continuous
everywhere (except, maybe, for the boundaries of different media), and try to find a reasonable partial differential equation for this $p(x,t)$.
The word ``reasonable'' refers to a requirement that the parameters (say, $D$) in this equation do not diverge or tend to zero identically almost everywhere.

Here we first show how the parametric scaling works for a homogeneous system: we take some $i$ and write (in time or in Laplace domain) for any function $f_i(t)$ or $f_i(s)$
\begin{eqnarray*}
 f_{i \pm 1}(z) &=& f(x,z) \pm a \frac{d}{dx} f(x,z) + \frac{a^2}{2} \frac{d^2}{dx^2} f(x,z) \\
 && \pm  \frac{a^3}{6} \frac{d^3}{dx^3} f(x,z) + ... \\
 &=& f(x,z) \pm a \frac{d}{dx} f(x,z) + \frac{a^2}{2} \frac{d^2}{dx^2} f(x,z) + O(a^3).
\end{eqnarray*}
with $z = t,s$, where $f(x,z)$ is the function interpolating $f_i(t)$ or $f_i(s)$ between the lattice points with $x_i = a i$. This means that e.g. in the Laplace domain
\begin{eqnarray}
 s p(x,s) - p(x,t=0) &&= \frac{a^2}{2} \frac{d^2}{dx^2} s M(x,s) p(x,s)  \nonumber \\
 &&+ \frac{a^3}{6} \frac{d^3}{dx^3} s M(x,s) p(x,s) + ... \; .
 \label{eq:Main_L}
\end{eqnarray}
For a homogeneous system all $M_i(s)$ are the same: $M_i(s)=M(s)$. Now we rescale $a$ and $\tau$, letting both tend to zero, and concentrate first on the second, temporal rescaling.

We consider the initial waiting time distributions in a form $\psi_i(t) = \tau^{-1} \phi(t/\tau)$ where $\tau$ represents a characteristic 
time scale for a jump, and discuss how the functions $s M(s) = s\psi(s)/[1-\psi(s)]$ are changed under taking the limit of $\tau \to 0$ for fixed $s$. 
As it follows from the scaling theorem, $f(bt) \leftrightarrow b^{-1} F(s/b)$. Thus
\begin{equation}
 \frac{1}{\tau} \phi \left(\frac{t}{\tau} \right) \leftrightarrow \phi(s \tau),
\end{equation}
and, for $s$ fixed  
\begin{equation}
sM(s)= \frac{ s\psi(s)}{1-\psi(s)} \to \frac{s \phi(s \tau)}{1 - \phi(s \tau)},
\end{equation}
where the argument of the Laplace-transformed $\psi$ gets automatically small for $\tau \to 0$.
We note that due to normalization $\lim_{s \to 0} \phi(s) = 1$, and therefore the corresponding function diverges for $\tau \to 0$. 
If the mean waiting time $\tau_0 = \langle t \rangle = c\tau$ exists (and is proportional to the characteristic time $\tau$, with the proportionality factor denoted by $c$), 
then $\psi(s \tau) \simeq 1 - c\tau s$. It gives us the asymptotic behavior
\begin{equation}
 sM(s) \simeq \frac{1}{c \tau}.
\end{equation}
If $\psi(t) \sim t^{-1-\alpha}$, the asymptotic behavior is 
\begin{equation}
 \frac{ s\psi(s)}{1-\psi(s)} \sim \frac{1}{\tau^\alpha} s^{1-\alpha}. 
 \label{eq:Ms}
\end{equation}
Note that the case for which the mean exists can be considered as a special case of the latter situation corresponding to $\alpha = 1$.
For the domains where all functions can be assumed as smooth we get the equation (for $0< \alpha < 1$)
\begin{eqnarray}
 sp_i(s) - p_i(t=0) &\simeq& \frac{a^2}{2 \tau^\alpha} \frac{d^2}{dx^2} s^{1-\alpha} p(x,s) \nonumber \\
 &&+ \frac{a^3}{6 \tau^\alpha} \frac{d^3}{dx^3} s^{1-\alpha} p(x,s) + ... \\
 &=& K_\alpha \frac{d^2}{dx^2} s^{1-\alpha} p(x,s) +  O \left( \frac{a^3}{\tau^\alpha} \right) \nonumber
 \label{eq:2}
\end{eqnarray}
with $K_\alpha = a^2/2\tau^\alpha$. Keeping this combination constant while taking $a \to 0$ and $\tau \to 0$ we see that the rest term vanishes (analog to Kramers-Moyal expansion)
while the first one on the r.h.s. stays finite. 
Any other scaling of $\tau$ w.r.t. $a$ makes no sense, since the r.h.s. of our equation either diverges (giving rise to the trivial limit of a solution vanishing
everywhere), or the coefficient in front of the lowest derivative vanishes together with all other coefficients (formally giving rise to the solution which does not depend on time).
Therefore, if the continuous description of our problem exists, it is given by an equation
\begin{eqnarray}
 sp(x,s) - p(x,t=0) &\simeq& \frac{a^2}{2 \tau^\alpha} \frac{d^2}{dx^2} s^{1-\alpha} p(x,s) \nonumber \\
 &=& K_\alpha  \frac{d^2}{dx^2} s^{1-\alpha} p(x,s),
 \label{fracdiffeq}
\end{eqnarray}
which, under passing to time domain, gets to be a fractional diffusion equation with a Riemann-Liouville derivative, 
\begin{equation}
 \frac{\partial}{\partial t} p(x,t) =  K_\alpha \frac{\partial^2}{\partial x^2} \;_0 D_t^{1-\alpha} p(x,t),
 \label{GDE}
\end{equation}
or, for $\alpha \to 1$,
a normal diffusion equation, the Fick's second law.

\subsection{Inhomogeneous parametric scaling limit}

We note that while in the case of a homogeneous system, when keeping the diffusion coefficient constant, it did not matter what indeed was rescaled, $a$ or $\tau$, in the case of the inhomogeneous system this matters. 

Let us consider an inhomogeneous system with $\alpha = \alpha(x)$ and $K_{\alpha(x)}(x)$ being slowly changing functions of $x$ (so that on the initial scale $\alpha^{-1}\frac{d \alpha}{dx} \ll a^{-1}$). 
If we consider some domain of the system in which $\alpha$ and $K$ may be considered as practically constant, the diffusion in
this domain will be described by the generalized diffusion equation, Eq.(\ref{GDE}), with the local coefficient of anomalous diffusion given by 
\begin{equation}
 K_{\alpha(x)}(x) = \frac{a^2}{2 \tau(x)^{\alpha(x)}}.
 \label{eq:K}
\end{equation}
This $K_{\alpha(x)}(x)$ is a physical characteristics of the domain, together with $\alpha(x)$. Both can be obtained by performing measurements at relatively short
times, when the particles do not leave the domain in which the parameters of diffusion do not change considerably. Now it is necessary to think how to take the continuous (van-Hove-like) limit correctly,
 i.e. whether to rescale $a$ or $\tau(x)$ by some factor $\lambda$ while keeping their combination, the local coefficient of anomalous diffusion, Eq.(\ref{eq:K}), constant.

If we are looking at a rescaling scheme which will keep all $a$'s the same, as it was in our initial system, i.e. take $a' = \lambda a$ corresponding to rescaling of all microscopic 
distances by a common scaling factor $\lambda$, the $\tau'(x) \to 0$ follows according to the corresponding restrictions on the local behavior,
\begin{equation}
 \tau'(x) = \lambda^{\frac{1}{\alpha(x)}} \tau(x) = \left[\lambda^2 \frac{a^2}{K_{\alpha(x)}(x)} \right]^{\frac{1}{\alpha(x)}},
  \label{eq:tau}
\end{equation}
so that all $\tau$ get an additional position dependence. 

Taking, on the contrary, all $\tau'(x) \to \lambda \tau(x)$ generates a spatially inhomogeneous lattice with $a'(x) = \sqrt{2 K_{\alpha(x)}(x) [\lambda \tau(x)]^{\alpha(x)}}$
which creates additional problems. A function simply interpolating between the values of $p_i(t)$ of the initial lattice does not have a probabilistic interpretation
(because there is a different number of sites of the new lattice between the two sites of the initial one), and the correction for this fact needs for a more complicated approach.
The situation for the normal diffusion on inhomogeneous lattices is considered in some detail in \cite{ItoStrat}, where it is shown that changes in $\tau$ or in $a$ essentially correspond to different 
interpretations of the limiting Langevin equations for the Brownian motion being the limit of the random walk scheme (namely to the \^{I}to and to the Stratonovich one, respectively) 
and therefore to different ensuing Fokker-Planck equations (with different solutions). These cannot hold true  simultaneously unless the spurious drift terms are correspondingly corrected for. Our initial scheme with symmetric jumps (constant $a$) corresponds to the \^{I}to case, and therefore the simplest approach is to keep this symmetry all the way on. 

Note that since $\alpha(x)$ is positive, all $\tau(x)$ tend to zero under fine-graining, and therefore all $\psi(s)$ tend to their small-$s$ asymptotic behavior.   
Let us now return to Eq.(\ref{eq:Main_L}), denote $F_i(s) = sM(x,s) p(x,s)$,
assume that $F_j$ are interpolated by some smooth function $F(x)$ so that $F(x) = F(a j)$, and perform the Taylor expansion:
\begin{equation}
\begin{split}{}
 sp_i(s) - p_i(t=0) \simeq a^2& \frac{d^2}{dx^2} \frac{s^{1-\alpha(x)}}{2 \tau(x)^{\alpha(x)}} p(x,s)\\ + & \tau(x)^{-\alpha(x)} O(a^3). 
 \end{split}
\end{equation}
Now we move the position-independent $a^2$ inside the derivative, use Eq.(\ref{eq:tau}) for $\tau(x)$ expressing this via $K_{\alpha(x)}$, and neglect the term of the higher order in $a$ (i.e. pass to the limit $a \to 0$):
\begin{equation}
 sp_i(s) - p_i(t=0) =  \frac{d^2}{dx^2} K_{\alpha(x)}(x) s^{1-\alpha(x)} p(x,s),
 \label{laplacediff}
\end{equation}
which in the time domain reads 
\begin{equation}
 \frac{\partial}{\partial t} p(x,t) = \frac{\partial^2}{\partial x^2} K_{\alpha(x)}(x) \;_0 D_t^{1-\alpha(x)}   p(x,t).
 \label{IGDE}
\end{equation}
Note that our parametric scaling scheme gives some physical flavor to the local scaling proposed by Straka \cite{straka2018variable}, in the sense that we now show that this scaling can be considered
as a typical condition that the continuous limit has to be taken such that the local coefficient of anomalous diffusion (assumed to exist) stays constant.

\section{Scaling Properties}
The discussions and simulations in Ref. \cite{straka2018variable} and in Ref. \cite{fedotov2019asymptotic} were performed for slightly different systems: in the first one the 
initial diffusion coefficient  $ K_{\alpha(x)}(x)$ was assumed constant, while in the second the original values of $\tau_i$ were taken constant: $\tau_i = \tau_0$,   which corresponds to $ K_{\alpha(x)}(x)$ changing with $x$. In simulations of \cite{fedotov2019asymptotic}
$\tau_0$ was chosen small ($\tau_0=10^{-3}$) and the diffusion coefficient was strongly position-dependent. Therefore, when 
comparing the results it is important to separate the two effects: The genuine effect of changes in $\alpha$, and the possible effect of strong inhomogeneity of the diffusion coefficients. 

In order to find out the relations between the corresponding solutions we first discuss what happens when we change the initial time and length scales of the system. 
\subsection{Temporal scaling}
First we look at the case where we choose $\tau_i = \tau_0$ constant. We start from equation (\ref{laplacediff}) and divide both parts of the equation by $s$ (in the time domain this corresponds
to integrating both parts of our integro-differential equation in time):
\begin{equation}
\begin{split}{}
 p(x,s) -\frac{p(x.t=0)}{s} =& \frac{\partial^{2}}{\partial x^2} \frac{a^2}{2 \tau_0^{\alpha(x)}} s^{-\alpha(x)} p(x,s)\\
 =& \frac{a^2}{2} \frac{\partial^{2}}{\partial x^2} (\tau_0 s)^{-\alpha(x)} p(x,s),
\end{split}
\end{equation}
Now we denote the new formal (dimensionless) Laplace frequency by $u=\tau_0 s$, express $s$ via $u$ and then divide both sides of the equation by $\tau_0$:
\begin{equation}{}
\frac{1}{\tau_0} p\left(x, \frac{u}{\tau_0} \right) - \frac{p(x.t=0)}{u} =  \frac{a^2}{2} \frac{\partial^{2}}{\partial x^2} u^{-\alpha(x)} \frac{1}{\tau_0} p\left(x, \frac{u}{\tau_0} \right).
\end{equation}
We introduce a new function 
\begin{equation}{}
g(x,u) = \frac{1}{\tau_0} p\left(x, \frac{u}{\tau_0} \right),
\end{equation}
and a new dimensionless time variable $\tilde{t} = t/\tau_0$. Remembering the previously mentioned scaling theorem, we see that the function $g$ is the (formal) Laplace transform of a function 
\begin{equation}
    G(x,\tilde{t}) = p(x,  \tau_0 \tilde{t})
\end{equation} 
in $\tilde{t}$.
Therefore by changing to the dimensionless time $\tilde{t} = t/\tau_0$, the initial equation, Eq. (\ref{laplacediff}), with $K_{\alpha(x)}(x) = a^2/2\tau_0^{\alpha(x)}$ can always be changed to an equation with position-independent effective diffusion coefficient $K_c = a^2/2$ (i.e. with the numeric value of $\tau_0$ equal to unity):
\begin{equation}
\frac{\partial}{\partial \tilde{t}} G(x,\tilde{t}) =  \frac{\partial^2}{\partial x^2}\;_0 D_{\tilde{t}}^{1-\alpha(x)} K_c G(x,\tilde{t}).
\end{equation}
This means that a system, in which $K_{\alpha(x)}$ depends on the position as $K_{\alpha(x)}=a^2/2\tau_0^{\alpha(x)}$, evolves $\tau_0^{-1}$ times faster (if $\tau_0 < 1$) or slower (if $\tau_0 > 1$) than the one in which $K_c$ was chosen constant, with $\tau_0=1$.\\
Now let us discuss the question, what happens if the diffusion coefficient is constant, but is not equal to $a^2/2$.

\subsection{Spacial scaling}
To relate the solution for the system with a constant diffusion coefficient $K_c$ but with $\tau_0 \neq 1$ to the
solution obtained in the previous case one has to spatially rescale the system. Let the numeric value of $K_c$ be $K_c = ca^2/2$.

Let us consider a system of size $L$ with the diffusion coefficient $K_c = ca^2/2$ whose time evolution is described 
by Eq. (\ref{laplacediff}), and change the lengthscale $x \to \tilde{x} = x/\sqrt{c}$. Under this change the length of the system changes to 
$\tilde{L} = L/\sqrt{c}$, and the behavior of $\alpha$ stays geometrically similar to the initial one: 
$\tilde{\alpha}(\tilde{x})=\alpha(x)$.  Now we  can rewrite equation  (\ref{laplacediff}) in the new variables:
\begin{eqnarray}
   sp_i(\tilde{x},s) - p_i(t=0) &=& \frac{1}{c} \frac{d^2}{d\tilde{x}^2} ca^2
 s^{1-\tilde{\alpha}(\tilde{x})} p(\tilde{x},s) \nonumber \\
 &=& \frac{d^2}{d\tilde{x}^2} a^2 s^{1-\tilde{\alpha}(\tilde{x})} p(\tilde{x},s).
 \label{eq:SpatRes}
\end{eqnarray}
The Dirichlet or Neumann boundary conditions are not influenced by the coordinate rescaling. 
Eq. (\ref{eq:SpatRes}) is of exactly the same form as Eq.  (\ref{laplacediff}) for $K_c = a^2/2$ and thus has the same solution. This means that the concentration profile at time $t$ in a system of size $L$ with $K_c = ca^2/2 $ is similar up to coordinate rescaling to the concentration profile in a system of size $\tilde{L} = L/\sqrt{c}$ with $K_c = a^2/2$ considered at the same time. This statement
allows for translation between the situations discussed in Refs. \cite{straka2018variable} and \cite{fedotov2019asymptotic}. 

\section{Numerical approach}
Now let us present the numerical procedure for solving the GME.
These solutions will be used as a benchmark for comparison with the
semi-analytical and with approximate solutions of the generalized diffusion
equations.

Starting from Eq.(\ref{eq:Main}), we may write 
\begin{equation}
\begin{split}
 -\frac{1}{2} sM_{i-1}(s)p_{i-1}(s) +\left[s+sM_{i}(s)\right]p_i(s) \\ - \frac{1}{2} sM_{i+1}(s)p_{i+1}(s) = p_i(t=0)
 \label{eq:intsite}
\end{split}
\end{equation}
for the internal sites of the system, and
\begin{equation}
\left[ s + \frac 1 2 sM_{1}(s)\right] p_1- \frac 1 2 sM_{2}(s) p_2 = p_1(t=0)
 \label{eq:endsite}
\end{equation}
for the sites at the boundaries of the interval. Eq. (\ref{eq:endsite})  follows from the choice of reflecting boundaries. For $i=[1,L]$ we consequently find a linear system of equations 
\begin{equation}
\mathbf{A}(s)\mathbf{ p}(s) = \mathbf{p}(t=0) 
\label{eq:matrix}
\end{equation}
with $\mathbf{A}(s)$ being a matrix of tridiagonal form depending on $s$ in the Laplace domain. Since the total probability within the system is conserved, the sum of all entries in a column of $\mathbf{A}(s)$ has to be  $s$. 
 
In the Laplace domain, the solution of Eq.(\ref{eq:matrix}) for the initial condition $\mathbf{p}(t = 0)$ is given by 
$ \mathbf{p}(s) = \mathbf{A}^{-1} (s)\mathbf{p}(t = 0)$. 
Having obtained this solution in the Laplace domain analytically or numerically, we have to perform its Laplace inversion to 
the time domain, pointwise in space. In the present work,  we invert the corresponding expressions or data numerically using  the Gaver-Stehfest algorithm \cite{gaver} which is best suited for finding inverse Laplace transforms of real functions whose originals 
do not oscillate. This is how the corresponding datapoints in Figs. \ref{result2dom} and \ref{solpics} are obtained.

\section{Example I: System with two domains}
\subsection{Matching conditions on the boundary of two media}

Let us now discuss the matching conditions of the probability density on the border of two different, homogeneous subdiffusive
media characterized by different parameters $\alpha$ and $K_\alpha$. The problem of a CTRW with different exponents $\alpha$
on the two sides of a border was posed in \cite{frac}. In this work the matching conditions for the case were simply guessed, 
and guessed wrongly. As mentioned in \cite{korabel2011boundary} the boundary conditions were corrected in \cite{marseguerra2006normal} and afterwards widely discussed \cite{shkilev2013boundary,kosztolowicz2018model,kosztolowicz2008subdiffusion,kosztolowicz2015random}.

\begin{figure}[ht]
\begin{center}
 \scalebox{0.50}{\includegraphics{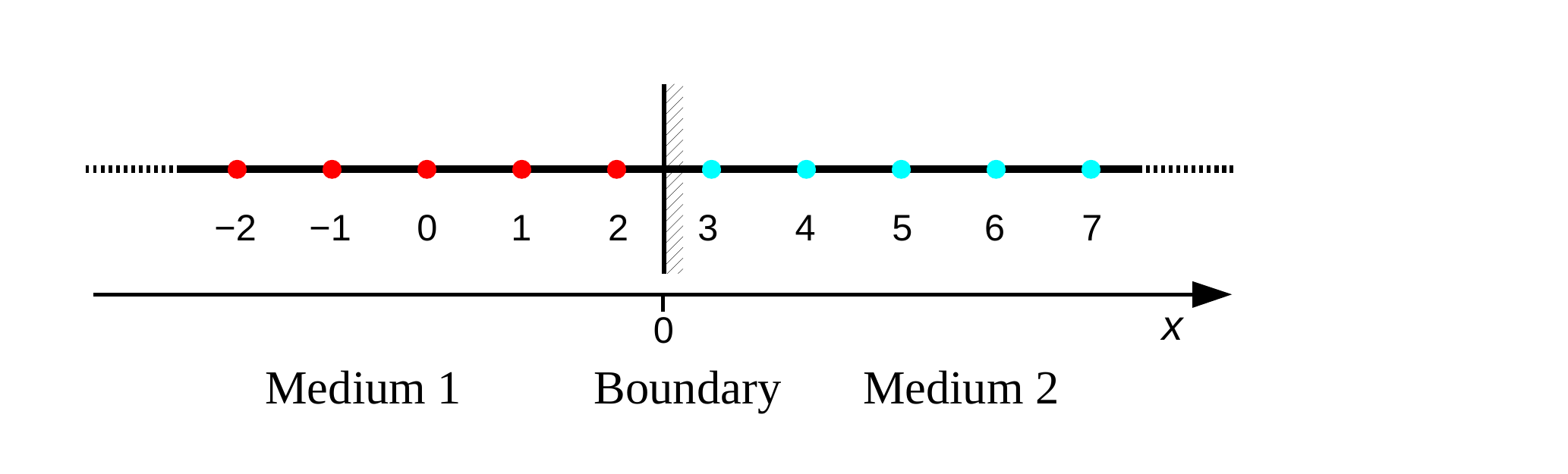}}
 \end {center}
\caption{Enumeration of sites at the boundary of two media as applied in Eq.(\ref{eq:sites}). \label{model}}
\label{modell}
\end{figure}

In order to find the
correct matching conditions we return to Eq. (\ref{gme}), and put down the corresponding equations explicitly for sites 2 and 3 immediately to the left and to the right of the boundary, see Fig. \ref{model}: 
\begin{eqnarray}
\frac{d}{dt} p_2(t) &=& \frac{1}{2}\hat{M}_{1}p_1(t)+\frac{1}{2}\hat{M}_{3}p_3(t) - \hat{M}_2p_2(t) , \nonumber \\
\frac{d}{dt} p_3(t) &=& \frac{1}{2}\hat{M}_{2}p_2(t)+\frac{1}{2}\hat{M}_{4}p_4(t) - \hat{M}_3p_3(t) . \label{eq:sites}
\end{eqnarray}
Here $\hat{M}_{1} = \hat{M}_{2} = \hat{M}_-$ correspond to the memory operators pertinent to medium 1, and $\hat{M}_{3} = \hat{M}_{4} = \hat{M}_+$ are the memory operators in medium 2. 
In the Laplace domain the equations read
\begin{equation}
    \begin{split}
         sp_2(s) - p_2(t_0) &= \frac{1}{2}sM_-p_1(s)+\frac{1}{2} sM_+p_3(s) - sM_-p_2(s) , \\  sp_3(s) - p_3(t_0) &= \frac{1}{2} sM_-p_2(s)+\frac{1}{2} sM_+p_4(s) - sM_+ p_3(s) .
    \end{split}
\end{equation}
Now we denote $p_2(s) = p_-(s)$, $p_3(s) = p_+(s)$ and approximate $p_1(s)$ and $p_4(s)$ via the derivatives of the continuous $p_{\pm}(x,s)$ right and left from the 
boundary. For example, to the right of the boundary we get: 
\begin{eqnarray}
&& sp_+(x,s) - p_+(x,0) = \label{eq:Border1} \\ 
&& \;\; \frac{1}{2}\left[sM_-p_--a\frac{d}{dx}sM_-p_-+ \frac{a^2}{2}\frac{d^2}{dx^2} sM_-p_- + O(M_-a^3) \right] \nonumber \\
&& \;\; +  \frac{1}{2}\left[sM_+p_+ +a\frac{d}{dx} sM_+p_+ + \frac{a^2}{2}\frac{d^2}{dx^2} sM_-p_+  + O(M_+a^3)\right] \nonumber \\
&& \;\;-sM_+p_+ \nonumber 
\end{eqnarray}
where on the right hand side of the equation all $p_\pm$ are to be understood as $p_{\pm}(x,s)$ and $M_{\pm}$ are $M_{\pm}(s)$. 
 
\begin{figure}[ht]
    {\includegraphics[width=0.5\textwidth]{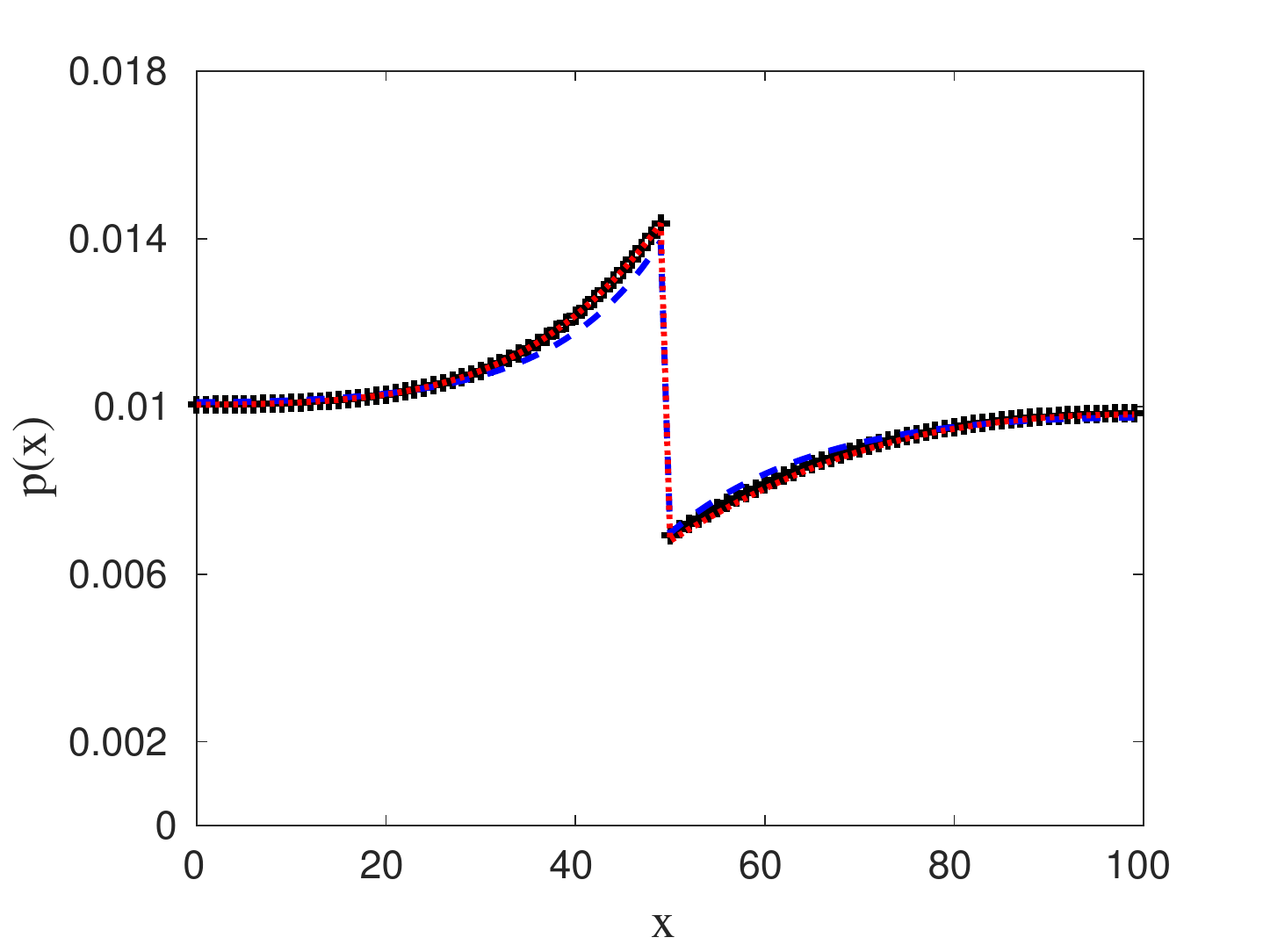}}
    {\includegraphics[width=0.5\textwidth]{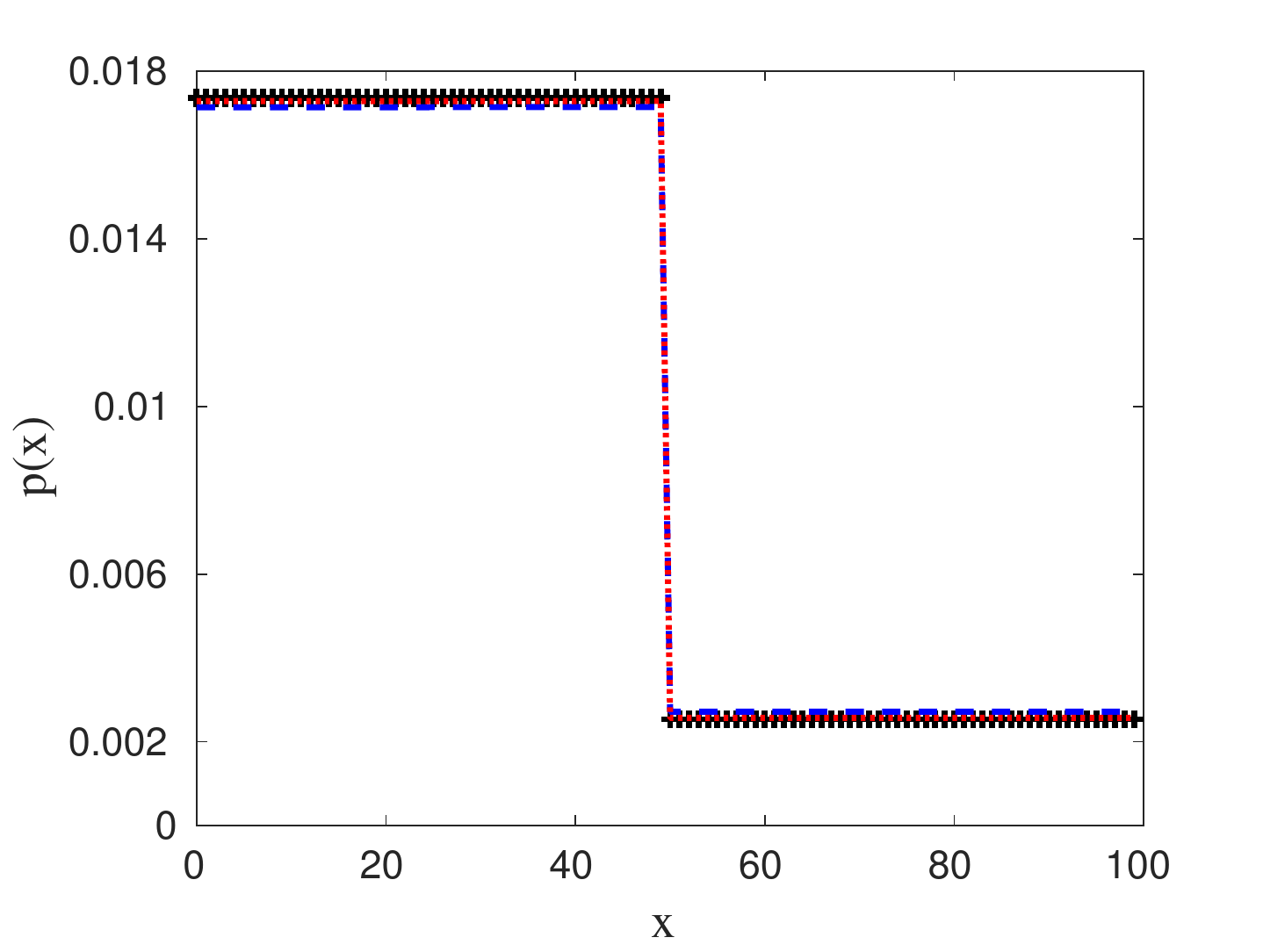}}
\caption{The PDF $p(x,t)$ in a system with $ \alpha_- = 0.8 $ in its left and $ \alpha_+ = 0.9$ in its right part
 for $t=10^{-2}$ (upper panel) and for $t=10^3$ (lower panel).
 The system consists of 100 sites, and the border is placed between sites 50 and 51. The initial condition is 
 $p_i(t=0) = 1/100$. The characteristic times are chosen with $\tau_{\pm} = \tau=10^{-5}$. 
 The graph shows the full numerical solution (black crosses), the (semi-)analytical  solution as obtained by numerical 
 Laplace inversion of Eq. (\ref{anasol2dom}) (red dotted line), and the approximation 
 as given by Eq.(\ref{tauberapprox}) (blue dashed line). }
\label{result2dom}
\end{figure}

Now we make a transition to the continuum taking $a \to 0$ as before, and keeping the diffusion coefficients $ K_{\alpha_\pm}
=a^2/\tau_{\pm}$ constant, so that $\tau_{\pm} \to 0$. The Laplace representations of the memory kernels then tend to 
\begin{equation}
 sM_{\pm}(s) \simeq \tau_{\pm}^{-\alpha_{\pm}} s^{1-\alpha_{\pm}},
\end{equation}
according to Eq.(\ref{eq:Ms}). Note that under the limiting transition the combinations $1/\tau_{\pm}$ and $ a/\tau_{\pm}$ diverge, 
$a^2/\tau_{\pm}$ stays constant, and $a^3/\tau_{\pm}$ and higher orders in $a$ vanish. We now assume that the solution for $p_{\pm}(x,s)$ exists and is smooth 
for $x>0$ and for $x<0$, with a possible singularity on the border. Under this assumption the diverging terms on the r.h.s. of Eq.(\ref{eq:Border1})
have to cancel, and we get:
\begin{equation}
\tau_{-}^{-\alpha_{-}} s^{1-\alpha_{-}} p_-(x,s) = \tau_{+}^{-\alpha_{+}} s^{1-\alpha_{+}} p_+(x,s)
\label{bound1}
\end{equation}
and
\begin{equation}
 \tau_{-}^{-\alpha_{-}} s^{1-\alpha_{-}} \frac{d}{dx} p_-(x,s) = \tau_{+}^{-\alpha_{+}} s^{1-\alpha_{+}} \frac{d}{dx} p_+(x,s),
 \label{bound2}
\end{equation}
i.e. the probability density and its first spacial derivative in the Laplace domain show a jump on the boundary.
These relations in the Laplace domain can be transformed to the time domain leading to integral relations for the probability densities and their derivatives on the boundaries. From Eq.(\ref{bound1}) it follows that 
\begin{equation}
\frac{p_+(s)}{p_- (s)} = \frac{ \tau_{-}^{-\alpha_{-}}}{\tau_{+}^{-\alpha_{+}}} s^{\alpha_+-\alpha_-},
\label{eq:quot}
\end{equation}
so that for $s\to 0$ $p_- (s)>> p_+(s)$ if $ \alpha_- < \alpha_+ $
(or vice versa, in the opposite case). This inequality in the Laplace domain is translated into a similar one in the time domain.
Therefore, at intermediate times, an inhomogeneous concentration profile establishes itself, leading to a 
particle flux from the domain with the larger value of $\alpha$ into the domain with the smaller value of
$\alpha$, and at longer times the whole probability concentrates in the domain with the smaller value of 
$\alpha$, as shown in Fig. \ref{result2dom}.

\subsection{Analytical solution in the Laplace domain} 
Eq. (\ref{laplacediff}) is an inhomogenous linear differential equation of second order.  For the case of two domains, its solutions can be easily found for each domain separately, and then matched. The corresponding general solutions read 
\begin{equation}
\begin{split}
    p_-(x,s) &=A_-e^{\gamma_- x}+B_-e^{-\gamma_- x}+\frac{p(t=0)}{s}, \\
    p_+(x,s) &=C_+e^{\gamma_+ x}+D_+e^{-\gamma_+ x}+\frac{p(t=0)}{s}
\end{split}
\end{equation}
with $\gamma_\pm(s) = \sqrt{2(s\tau)^{\alpha_\pm}}/a_0$, and with four
integration constants $A_-, B_-, C_+$ and $D_+$. The boundary conditions at the outer boundaries of the system, which are taken 
to be reflecting (no-flux) ones, $\partial_x p_-\big \vert _{x=0} = 0$ and $\partial_x p_+\big \vert _{x=L} = 0$, fix two of the integration constants.  With 
these conditions we obtain the following expressions for $p_\pm(x,s)$: 
\begin{equation}
\label{anasol2dom}
\begin{split}
    p_-(x,s) &=2A_-\cosh(\gamma_- x)+\frac{p(t=0)}{s}, \\
    p_+(x,s) &=C_+(e^{\gamma_+ x}+e^{2\gamma_+L}e^{-\gamma_+ x})+\frac{p(t=0)}{s} .
\end{split}
\end{equation}
The integration constants  $C_+$ and $A_-$ follow then from the matching conditions, Eqs. (\ref{bound1}) and (\ref{bound2}) and read
\begin{equation}
    \begin{split}
        C_+ =& \frac{(\Theta-1)\tanh\left(\gamma_-\frac{L}{2}\right)p(t=0)}{ \Theta s}\\
        & \times \Bigg[(e^{\gamma_+\frac{L}{2}}-e^{2\gamma_+L}e^{-\gamma_+\frac{l}{2}})\frac{\gamma_+}{\gamma_-} \\ &-\tanh\left(\gamma_- \frac{L}{2}\right)(e^{\gamma_+\frac{L}{2}}+e^{2\gamma_+L}e^{-\gamma_+\frac{l}{2}})\bigg]^{-1}
    \end{split}
\end{equation}
and
\[
A_- =  C_+ \Theta  \frac{\gamma_+}{\gamma_-}\frac{(e^{\gamma_+\frac{L}{2}}-e^{2\gamma_+L}e^{-\gamma_+\frac{l}{2}})}{2\sinh(\gamma_-\frac{L}{2})} 
\] 
with $\Theta(s) = p_+(s)/p_- (s)$ as given by Eq. (\ref{eq:quot}).
Having the analytical solution in the Laplace domain we can numerically invert it to the time domain. As evident from  Fig. \ref{result2dom}, this semi-analytical solution can hardly be distinguished from the full numerical solution of a discrete system
(i.e. the solution of Eq.(\ref{eq:matrix}) with the subsequent Laplace inversion).
We can also give simple analytical estimates of how the system behaves for very short and for very long times. 
Assuming that in the time domain the behavior of $p(x,t)$ for fixed $x$ is a power law, possibly modulated by some 
slowly varying function, one can perform the Laplace inversion for very short and for very long times approximately, applying a Tauberian theorem:  
\begin{equation}
\begin{split}
    f(t)  &\cong t^{\rho-1} L(t) \\
    &\Updownarrow \\
    f(s)  &\cong \Gamma(\rho) s^{-\rho} L(1/s)
\end{split}
\end{equation}
with $L(t)$ being a slowly varying function of its argument. Therefore we my write
\begin{equation}
    p(x,t) \approx \frac{1}{t} p \left(x,s=\frac{1}{t}\right)
    \label{tauberapprox}
\end{equation}
(since we don't know $\rho$ exactly, we simply assume that $1/\Gamma(\rho)$ is of the order of unity, and omit
this prefactor).  As it can be seen in Fig. \ref{result2dom} and \ref{solpics} for the cases of the two domains with constant 
$\alpha(x)$, and of linearly changing $\alpha(x)$ (as discussed in the next section) respectively, 
the approximation, Eq. (\ref{tauberapprox}), performs for large times very well, and for small times has a relative 
accuracy of the order of 10\%.

\section{Example II : Linear change in $\alpha(x)$}
 \begin{figure}[ht]
     \begin{center}
     {\includegraphics[width=0.5\textwidth]{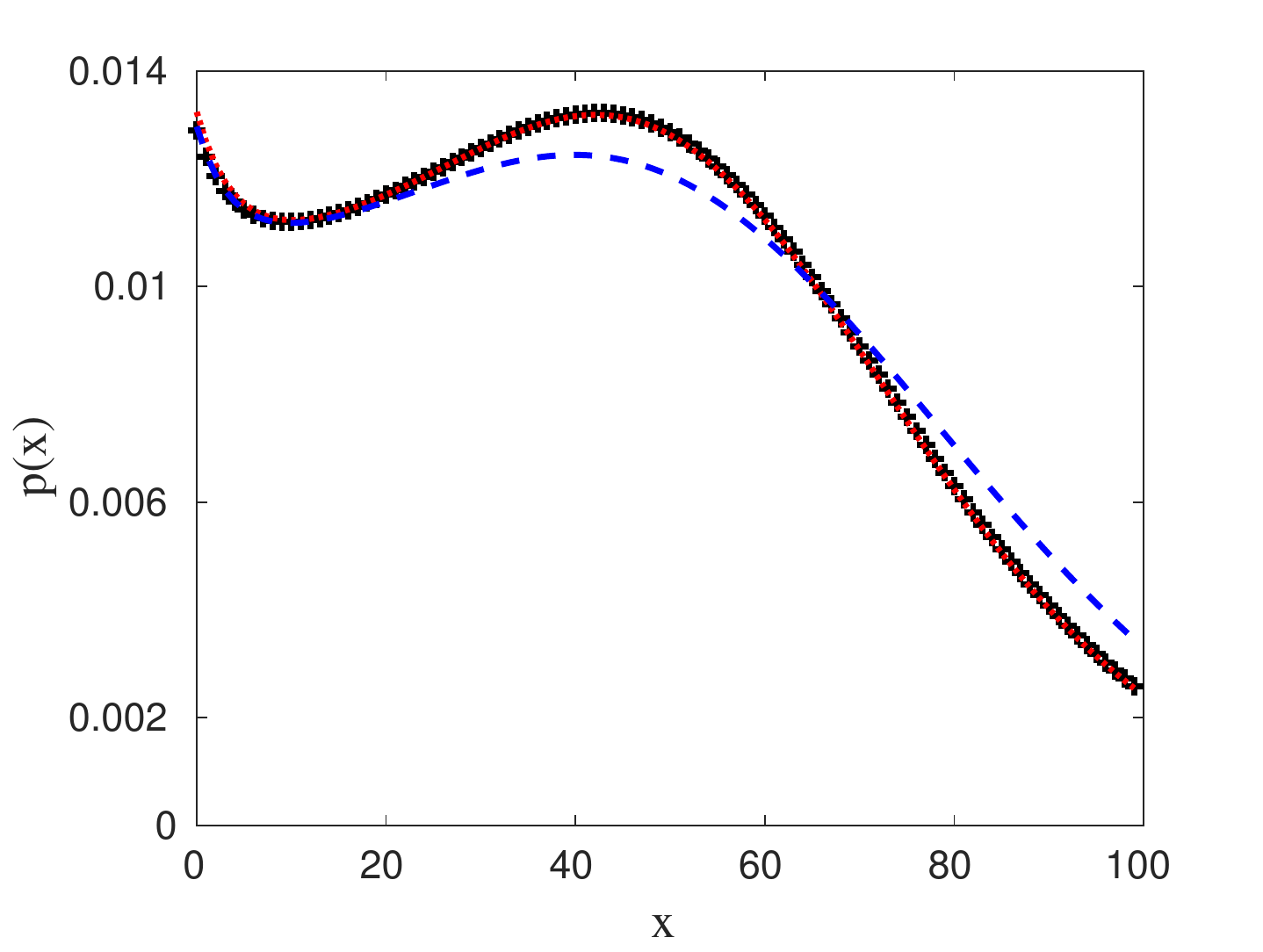}}
     {\includegraphics[width=0.5\textwidth]{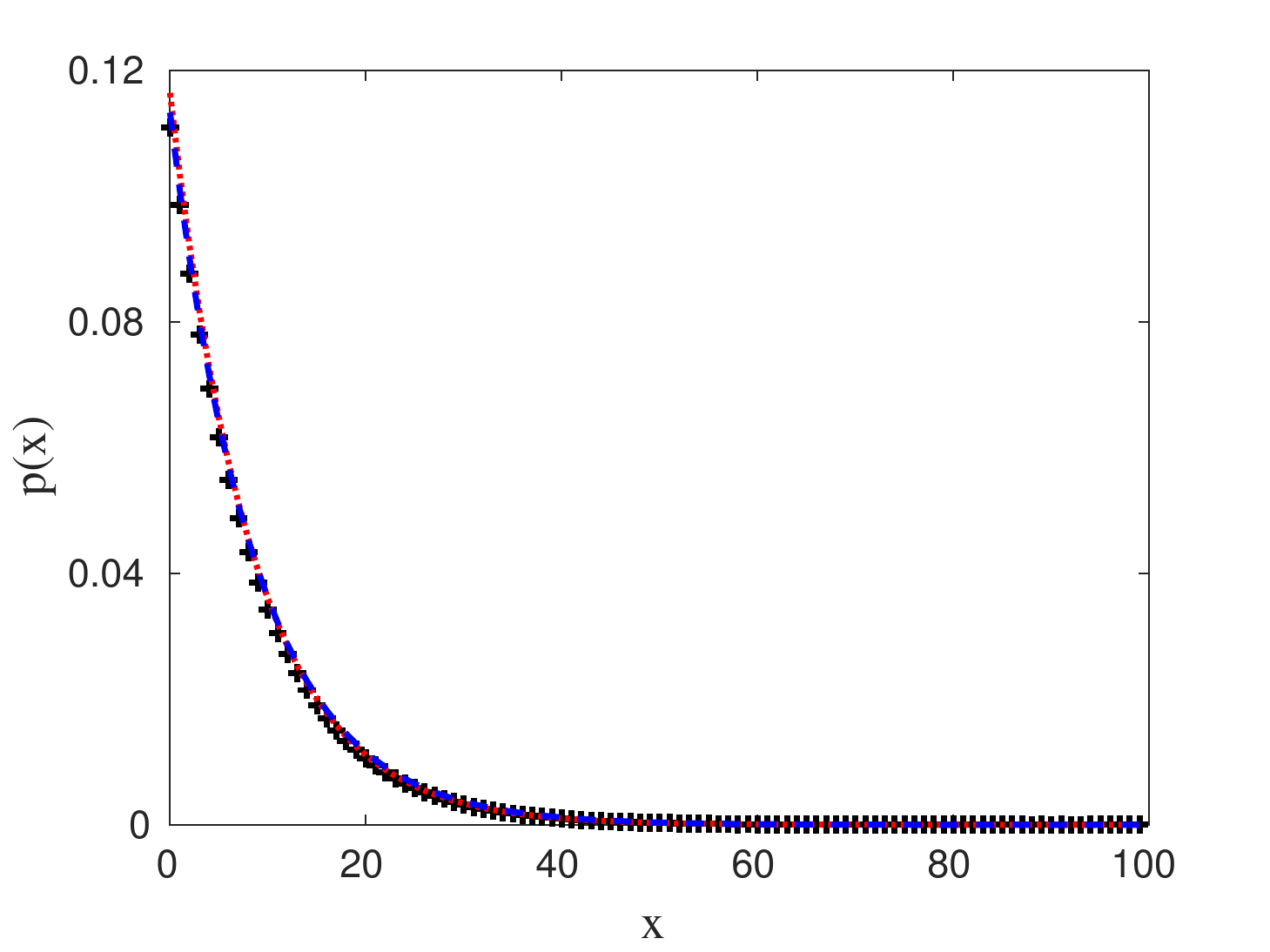}}
     \end{center}
    \caption{The PDF $p(x,t)$ in a system with linearly changing $\alpha(x)$ for time $t=0.1$ (upper panel) and $t=10^5$ (lower panel). We used $\alpha(x)= 0.4+0.005x$ for $x=[1,100]$. Choosing $\tau=10^{-5}$ led to $K_{\alpha(0)}=100$ on the left boundary and $K_{\alpha(L)}\approx3\cdot10^4$ on the right boundary of the interval. The initial condition is $p(x,t=0)=1/100$. The results of the numerical solution of the GME are shown as crosses, the (semi-)analytical solution obtained by a numerical Laplace inversion of the analytical solution, Eq.(\ref{eq:fin_lin}) are shown as red dotted lines, and the results of approximate analytical inversion, Eq.(\ref{tauberapprox}), by a blue dashed ones.}
     \label{solpics}
 \end{figure}
 
  Now we turn to another example, the system with the linear change in $\alpha(x)$ as discussed in  \cite{fedotov2019asymptotic}.
 Starting from the equation in the Laplace domain, Eq. (\ref{fracdiffeq}), 
and introducing a new dependent variable $F_s(x)$ defined according to
\begin{equation}
 p(x,s) = \frac{s^{\alpha(x)-1}}{K_{\alpha(x)}} F_s(x),
\end{equation}
we obtain for this new variable the equation 
\begin{equation}
   \frac{s^{\alpha(x)}}{K_{\alpha(x)}} F_s(x) -p(t=0)= \frac{\partial^2}{\partial x^2} F_s(x),
   \label{frac2}
\end{equation}
a linear inhomogeneous second-order differential equation with a constant coefficient in front of the second derivative. 
We now look at the case where the diffusion exponent grows linear with spatial coordinate: $\alpha(x) = c+bx $.  Assuming $\tau$ to be constant, and therefore taking $K_{\alpha(x)} = a^2/2 \tau^{\alpha(x)}$, we may write:
\begin{equation}
    \frac{\partial^2}{\partial x^2} F_s(x) - \omega e^{\epsilon  x} F_s(x) + p(t=0)= 0
    \label{orig}
\end{equation}
with $\omega  = 2 (s\tau)^c/a_0^2$ and $\epsilon  = \ln(s\tau)b$. Eq. (\ref{orig}) can be put into a form of an inhomogeneous Bessel equation. There exist several variable transformations to get to the Bessel equation, two of which are presented in appendix \ref{sec:appa}. The necessary to present two different variable transformations is connected with the 
necessity to show that the solutions of Eq. (\ref{frac2}) in the corresponding domains are real. 

Fist we substitute $ \zeta =  2\sqrt{\omega }\epsilon ^{-1} e^{\epsilon  x/2} $ and define $F_s(x) = g(\zeta)$. 
The second spacial derivative becomes now
\begin{equation}
    \frac{\partial^2 F_s(x)}{\partial x^2} = \frac{\partial^2 g(\zeta)}{\partial \zeta^2} \left( \frac{\partial \zeta(x)}{\partial x}\right)^2 + \frac{\partial g(\zeta)}{\partial x} \frac{\partial^2 \zeta(x)}{\partial x^2}.
\end{equation}
Plugging this expression into our original equation, Eq. (\ref{orig}), with $\partial_x \zeta(x) =  \sqrt{\omega }e^{\epsilon  x/2}$ and $\partial_x^2 \zeta(x) = \frac{\epsilon }{2}  \sqrt{\omega }e^{\epsilon  x/2}$, we get the inhomogeneous 
Bessel equation:
\begin{equation}
    \label{bessel}
    \zeta^2 \frac{\partial ^2 g(\zeta)}{\partial \zeta^2} + \zeta \frac{\partial g(\zeta)}{\partial \zeta} - \zeta^2 g(\zeta) = - \frac{4}{\epsilon ^2}p(t=0) .
\end{equation}
The solution to the homogeneous part of this equation is a linear combination of the modified Bessel functions $I_0$ and $K_0$. A particular solution of the inhomogeneous equation is the sum of products of $I_0$ and $K_0$ with Meijer G-functions. Thus, solving Eq.(\ref{fracdiffeq}) for a linearly growing $\alpha(x)$ and reflecting boundary conditions it we get: 
\begin{eqnarray}
 &&  p(x,s) = \frac{2(s\tau)^{\alpha(x)}}{sa_0^2}
 \Bigg\{C_1 I_0(\zeta) + C_2 K_0 (\zeta) +\frac{p(t=0)}{\epsilon ^2}  \\ 
 &&\times \left. \bigg[  I_0(\zeta)\MeijerG{0}{1}{3}{3}{1}{0,0,0}{\frac{\zeta^2}{4}} -2K_0(\zeta) \MeijerG{0}{1}{2}{3}{1}{0,0,0}{\frac{-\zeta^2}{4}}\bigg] \right\} 
 \nonumber 
\label{eq:fin_lin}
\end{eqnarray}
The constants $C_1$ and $C_2$, and the procedure of finding the particular solution can also be found in Appendix \ref{sec:appa}. 
The Meijer G-Function $\MeijerG{m}{n}{p}{q}{a_1,\ldots,a_p}{b_1,\ldots,b_q}{\zeta}$ is defined as 
\begin{equation}
\begin{split}{}
    &\MeijerG{m}{n}{p}{q}{a_1,\ldots,a_p}{b_1,\ldots,b_q}{x} = \\ &\frac{1}{2 \pi i} \int_{\gamma L}\frac{\Pi_{j=1}^m \Gamma(b_j+s)\Pi_{j=1}^n \Gamma(1-a_j+s)}{\Pi_{j=n+1}^p \Gamma(a_j+s)\Pi_{j=m+1}^q \Gamma(1-b_j+s)}x^{-s}ds,
\end{split}
\end{equation}
(see e.g. p. 793 of Ref. \cite{prudnikovyu}).

The solution of Eq. (\ref{frac2}) has to be real, since its multiplication
by another real function should give the Laplace transform of the PDF.
The full solution for the PDF, Eq. (\ref{eq:fin_lin}), is evidenltly real for $\zeta > 0$, when both Bessel functions and the Meijer  G-Function of a positive argument are real. However for $\zeta < 0$ (i.e. for $\tau s < 1$) the function $K_0(\zeta)$ of a negative argument may acquire an imaginary part. Since the overall structure of the solution, Eq. (\ref{eq:fin_lin}), involving integration constants given by quite complex  expressions, is quite awkward for the analysis, it is not immediately obvious that the solution Eq. (\ref{eq:fin_lin}) stays real also for $\zeta < 0$. Here, another change of variables outlined in Appendix \ref{sec:appa},
shows that the solution stays real also for $\zeta < 0$. Since the solution of Eq. (\ref{frac2}) is unique, these two different solutions have to 
match for $\zeta \to 0 $. We didn't check this analytically, but from the numerics show that they do.  
The numerical inverse Laplace transform of the analytical solution, Eq.(\ref{eq:fin_lin}), was again performed by using the Gaver-Stehfest algorithm. 
The solution of the PDF for a continuous approximation was then compared to the full numerical solution for a discrete 
system. The results of such a comparison are presented in Fig. \ref{solpics} together with the analytical approximation 
as given by Eq. (\ref{tauberapprox}), showing that the solution of the 
continuous equation indeed excellently reproduces the numerical solution
of the GME.
\label{sec:lin}

\section{Conclusions}

In this article, we discuss the continuous limit of a lattice CTRW scheme with power-law WTDs at lattice sites, with position-dependent parameters. The parametric scaling of local parameters in the WTD leads to a variable-order time-fractional diffusion equation. Different choices of local parameters of the WTDs change the behavior of the subdiffusion coefficient of the system, but the solutions of the corresponding equations can be related to each other in terms of their temporal and spacial rescaling. 

As examples we discussed two different situations. The first one corresponds to a  system with a border separating two different media, with different exponents of the power-law WTDs. For this case we derive the matching conditions of the solutions on that border. As a second example we study a system where the diffusion exponent changes linearly with the position. 

For both examples we compare numerical solutions of the initial GMEs with the ones of the variable-order diffusion equations, and show that they perfectly match. Both solutions are first obtained in the Laplace domain
(numerically for GME, analytically for its continuous counterpart) and than transformed numerically to the time domain with the help of the Gaver-Stehfest algorithm. We moreover show that approximate analytical solutions of the continuous problem can be obtained for short and for long times by applying a Tauberian theorem. 

\appendix

\section{}
\label{sec:appa}

In this Appendix we first discuss another variable transformation reducing
Eq.(\ref{frac2}) to a Bessel equation, and then discuss the form of the general solution and the values of integration constants in Eq.(\ref{eq:fin_lin}) of the main text. 

In section \ref{sec:lin} we mentioned that the total solution of Eq. (\ref{frac2}) has to be real. For the presented solution, Eq. (\ref{eq:fin_lin}), this is only evident if $\tau s > 1$ ($ \zeta > 0$) because only then the modified Bessel functions stay real. For $\tau s < 1$ ($\zeta < 0 $)  $K_0$ of a negative argument is in general complex. This is why we look for another  variable transformation $x \to \tilde{\zeta}$ so that $\tilde{\zeta}$ stays positive in the domain $\tau s < 1$. 
To this end we look at the system mirrored with respect to the middle of the interval. Equation (\ref{frac2}) then reads  
\begin{equation}
\begin{split}{}
   \frac{s^{\tilde \alpha(y)}}{K_{\tilde \alpha(y)}} \tilde F_s(y) -p(t=0)= \frac{\partial^2}{\partial y^2} \tilde F_s(y)
    \end{split}
    \label{eq:app}
\end{equation}
with the substitution $y = L-x$ and $\tilde{F}_s(y) = F_s(x)$. 
The equation is of the same form as Eq. (\ref{frac2}) but with \begin{equation}
  \tilde \alpha(y) =\alpha(x) = c^*+b^*y  
\end{equation}
with $c^* = c+bL$ and $b^* = -b$. We can now introduce a new independent variable 
\begin{equation}{}
\tilde \zeta(y) = 2\sqrt{\frac{2(s \tau)^{c^*}}{a_0^2}}\frac{1}{\ln{(s\tau)}}e^{-\ln{(s \tau)}y/2} 
\end{equation}
to get to the Bessel equation. The variable changes from $y$ resp. $x$ to $\tilde{\zeta}(y)$ and $\zeta(x)$ are of the same form but with different parameters. The new independent variable $\tilde{\zeta}(y)$ is positive for $\tau s < 1 $ and thus $\tilde{F}_s(y)$ for this case is evidently real. Since $\tilde{F}_s(y) = F_s(x)$ the total solution is real for both cases $\tau s < 1 $  and $\tau s > 1 $.

Now we discuss the procedure of solving Eq. (\ref{bessel}) and therefore also Eq. (\ref{eq:app}) since they only differ in their
parameter values. 

The solution of Eq. (\ref{bessel}) is a sum of a particular solution of
the inhomogeneous equation and of a general solution of the 
corresponding homogeneous one. The solution of the homogeneous equation is a linear combination of the modified Bessel functions of first and second kind ($I_0$ and $K_0$):
\begin{equation}
\begin{split}
       g(\zeta) &= g_{p}(\zeta)+g_{hom}(\zeta)  \\
                    &= g_{p}(\zeta)+C_1I_0(\zeta) + C_2K_0(\zeta).
\end{split}
\end{equation}
To find the particular solution for the inhomogeneous equation we use the ansatz $g_{p}(\zeta)=A_1(\zeta)I_0(\zeta)+A_2(\zeta) K_0(\zeta)$  (variation of the constants). The derivatives of $A_1(\zeta)$ and $A_2(\zeta)$ are given by:
\begin{equation}
\begin{split}
      A_1'(\zeta) = \frac{K_0(\zeta)}{W(\zeta)}\frac{4p(t=0)}{\epsilon ^2 \zeta^2} \\
      A_2'(\zeta) = -\frac{I_0(\zeta)}{W(\zeta)}\frac{4p(t=0)}{\epsilon ^2 \zeta^2}
\end{split}
\end{equation}
with $W(x)=I_0(\zeta)K_1(\zeta)-I_1(\zeta)K_0(\zeta)$ being the Wronsky determinant. Integrating these terms with MATHEMATICA we find the complete form of the particular solution:
\begin{equation}
\begin{split}
    &g_{p}(\zeta) = \frac{p(t=0)}{\epsilon ^2} \times \\ &\left[ I_0(\zeta)\MeijerG{0}{1}{3}{3}{1}{0,0,0}{\frac{\zeta^2}{4}} -2K_0(\zeta) \MeijerG{0}{1}{2}{3}{1}{0,0,0}{\frac{-\zeta^2}{4}} \right].
\end{split}
\end{equation}
For getting the complete solution we still have to determine the integration constants $C_1$ and $C_2$ which
follow from the no-flux boundary conditions at $x=0$ and $x=L$. Thus: 
\begin{equation}
\label{bound}
\begin{split}{}
J \bigg \vert _{x=0,L}&= \frac{\partial}{\partial x}\left[K_{\alpha(x)} s^{1-\alpha(x)} p(x,s) \right] _{x=0,L}\\  &= \frac{\partial}{\partial x} F_s(x) \bigg \vert _{x=0,L}= 0.
\end{split}
\end{equation}
Since $\partial_x F_s = \frac{g(\zeta)}{\partial{\zeta}} \frac{\partial \zeta}{\partial x}$
and $\partial_x \zeta(x) =  \sqrt{\omega }e^{\epsilon  x/2} \neq 0 $ we can put:
\begin{equation}
     \frac{g(\zeta)}{\partial{\zeta}}\bigg \vert _{x=0,L}= 0.
\end{equation}
To fulfill these condition we need to set the constants as follows: 
\begin{widetext}
\begin{eqnarray*}
        C_1 &=& \frac{- p(t=0)}{ \lambda^2(I_1(\zeta_L)K_1(\zeta_0)-I_1(\zeta_0)K_1(\zeta_L))} 
       \left[ 2K_1(\zeta_0)K_1(\zeta_L) G_2\left(\frac{-\zeta_0^2}{4}\right) \right. \\ 
       && \left. +2K_1(\zeta_0)K_1(\zeta_L)G_2\left(\frac{-\zeta_L^2}{4}\right)
       -I_1(\zeta_0)K_1(\zeta_L)G_1\left(\frac{\zeta_0^2}{4}\right)-I_1(\zeta_L)K_1(\zeta_0)G_1\left(\frac{\zeta_L^2}{4}\right)\right],
\end{eqnarray*}

\begin{eqnarray*}
       C_2 &= &\frac{- p(t=0)}{ \lambda^2(I_1(\zeta_L)K_1(\zeta_0)-I_1(\zeta_0)K_1(\zeta_L))} 
 \left[ 2I_1(\zeta_L)K_1(\zeta_0) G_2\left(\frac{-\zeta_0^2}{4}\right) \right. \\
 	&& \left. +2I_1(\zeta_0)K_1(\zeta_L)G_2\left(\frac{-\zeta_L^2}{4}\right) -I_1(\zeta_0)I_1(\zeta_L)G_1\left(\frac{\zeta_0^2}{4}\right)-I_1(\zeta_0)I_1(\zeta_L)G_1\left(\frac{\zeta_L^2}{4}\right)\right]
\end{eqnarray*}
\end{widetext}
where the following abbreviations were used:
\begin{eqnarray*}
       G_1(x) &=&   \MeijerG{0}{1}{3}{3}{1}{0,0,0}{x} ,\\
       G_2(x) &=& \MeijerG{0}{1}{2}{3}{1}{0,0,0}{x} ,\\
       \zeta_0  &=&  \frac{2\sqrt{\omega }}{\epsilon } ,\\
       \zeta_L  &=&  \frac{2\sqrt{\omega }}{\epsilon } e^{\epsilon  L/2} .
\end{eqnarray*}
\newline

Going back to $F_s(x)$ we can write down: 
\begin{equation}
\label{solu}
    F_s(x) = C_1 I_0 \left(\frac{2\sqrt{\omega }}{\epsilon } e^{\epsilon  x/2} \right) + C_2 K_0 \left(\frac{2\sqrt{\omega }}{\epsilon } e^{\epsilon  x/2} \right) + F_s^{p} ,
\end{equation}
and thus for the PDF:
\begin{equation}
\label{finalsol}
 p(x,s) = \frac{2(s\tau)^{\alpha(x)}}{s a_0^2} F_s(x) .
\end{equation}


\begin{thebibliography}{24}%
\makeatletter
\providecommand \@ifxundefined [1]{%
 \@ifx{#1\undefined}
}%
\providecommand \@ifnum [1]{%
 \ifnum #1\expandafter \@firstoftwo
 \else \expandafter \@secondoftwo
 \fi
}%
\providecommand \@ifx [1]{%
 \ifx #1\expandafter \@firstoftwo
 \else \expandafter \@secondoftwo
 \fi
}%
\providecommand \natexlab [1]{#1}%
\providecommand \enquote  [1]{``#1''}%
\providecommand \bibnamefont  [1]{#1}%
\providecommand \bibfnamefont [1]{#1}%
\providecommand \citenamefont [1]{#1}%
\providecommand \href@noop [0]{\@secondoftwo}%
\providecommand \href [0]{\begingroup \@sanitize@url \@href}%
\providecommand \@href[1]{\@@startlink{#1}\@@href}%
\providecommand \@@href[1]{\endgroup#1\@@endlink}%
\providecommand \@sanitize@url [0]{\catcode `\\12\catcode `\$12\catcode
  `\&12\catcode `\#12\catcode `\^12\catcode `\_12\catcode `\%12\relax}%
\providecommand \@@startlink[1]{}%
\providecommand \@@endlink[0]{}%
\providecommand \url  [0]{\begingroup\@sanitize@url \@url }%
\providecommand \@url [1]{\endgroup\@href {#1}{\urlprefix }}%
\providecommand \urlprefix  [0]{URL }%
\providecommand \Eprint [0]{\href }%
\providecommand \doibase [0]{http://dx.doi.org/}%
\providecommand \selectlanguage [0]{\@gobble}%
\providecommand \bibinfo  [0]{\@secondoftwo}%
\providecommand \bibfield  [0]{\@secondoftwo}%
\providecommand \translation [1]{[#1]}%
\providecommand \BibitemOpen [0]{}%
\providecommand \bibitemStop [0]{}%
\providecommand \bibitemNoStop [0]{.\EOS\space}%
\providecommand \EOS [0]{\spacefactor3000\relax}%
\providecommand \BibitemShut  [1]{\csname bibitem#1\endcsname}%
\let\auto@bib@innerbib\@empty
\bibitem [{\citenamefont {Metzler}\ and\ \citenamefont
  {Klafter}(2000)}]{MetzKla}%
  \BibitemOpen
  \bibfield  {author} {\bibinfo {author} {\bibfnamefont {R.}~\bibnamefont
  {Metzler}}\ and\ \bibinfo {author} {\bibfnamefont {J.}~\bibnamefont
  {Klafter}},\ }\bibfield  {title} {\enquote {\bibinfo {title} {The random
  walk's guide to anomalous diffusion: a fractional dynamics approach},}\
  }\href@noop {} {\bibfield  {journal} {\bibinfo  {journal} {Phys. Rep.}\
  }\textbf {\bibinfo {volume} {339}},\ \bibinfo {pages} {1} (\bibinfo {year}
  {2000})}\BibitemShut {NoStop}%
\bibitem [{\citenamefont {Sokolov}(2012)}]{SoftMatt}%
  \BibitemOpen
  \bibfield  {author} {\bibinfo {author} {\bibfnamefont {I.M.}\ \bibnamefont
  {Sokolov}},\ }\bibfield  {title} {\enquote {\bibinfo {title} {Models of
  anomalous diffusion in crowded environments},}\ }\href@noop {} {\bibfield
  {journal} {\bibinfo  {journal} {Soft Matter}\ }\textbf {\bibinfo {volume}
  {8}},\ \bibinfo {pages} {9043} (\bibinfo {year} {2012})}\BibitemShut
  {NoStop}%
\bibitem [{\citenamefont {Liang}\ \emph {et~al.}(2019)\citenamefont {Liang},
  \citenamefont {Wang}, \citenamefont {Chen}, \citenamefont {Zhou},\ and\
  \citenamefont {Magin}}]{Liang}%
  \BibitemOpen
  \bibfield  {author} {\bibinfo {author} {\bibfnamefont {Y.}~\bibnamefont
  {Liang}}, \bibinfo {author} {\bibfnamefont {S.}~\bibnamefont {Wang}},
  \bibinfo {author} {\bibfnamefont {W.}~\bibnamefont {Chen}}, \bibinfo {author}
  {\bibfnamefont {Zh.}\ \bibnamefont {Zhou}}, \ and\ \bibinfo {author}
  {\bibfnamefont {R.L.}\ \bibnamefont {Magin}},\ }\bibfield  {title} {\enquote
  {\bibinfo {title} {A survey of models of ultraslow diffusion in heterogeneous
  materials},}\ }\href@noop {} {\bibfield  {journal} {\bibinfo  {journal}
  {Appl. Mech. Rev.}\ }\textbf {\bibinfo {volume} {71(2)}},\ \bibinfo {pages}
  {040802} (\bibinfo {year} {2019})}\BibitemShut {NoStop}%
\bibitem [{\citenamefont {Klages}\ \emph {et~al.}(2008)\citenamefont {Klages},
  \citenamefont {Radons},\ and\ \citenamefont {Sokolov}}]{klagessok}%
  \BibitemOpen
  \bibfield  {author} {\bibinfo {author} {\bibfnamefont {R.}~\bibnamefont
  {Klages}}, \bibinfo {author} {\bibfnamefont {G.}~\bibnamefont {Radons}}, \
  and\ \bibinfo {author} {\bibfnamefont {I.M.}\ \bibnamefont {Sokolov}},\
  }\href@noop {} {\emph {\bibinfo {title} {Anomalous transport: foundations and
  applications}}}\ (\bibinfo  {publisher} {Wiley-VCH, Weinheim},\ \bibinfo
  {year} {2008})\BibitemShut {NoStop}%
\bibitem [{\citenamefont {Chechkin}\ \emph {et~al.}(2005)\citenamefont
  {Chechkin}, \citenamefont {Gorenflo},\ and\ \citenamefont {Sokolov}}]{frac}%
  \BibitemOpen
  \bibfield  {author} {\bibinfo {author} {\bibfnamefont {A.V.}\ \bibnamefont
  {Chechkin}}, \bibinfo {author} {\bibfnamefont {R.}~\bibnamefont {Gorenflo}},
  \ and\ \bibinfo {author} {\bibfnamefont {I.M.}\ \bibnamefont {Sokolov}},\
  }\bibfield  {title} {\enquote {\bibinfo {title} {Fractional diffusion in
  inhomegenous media},}\ }\href@noop {} {\bibfield  {journal} {\bibinfo
  {journal} {J. Phys. A}\ }\textbf {\bibinfo {volume} {38}},\ \bibinfo {pages}
  {L679} (\bibinfo {year} {2005})}\BibitemShut {NoStop}%
\bibitem [{\citenamefont {Korabel}\ and\ \citenamefont {Barkai}(2010)}]{para}%
  \BibitemOpen
  \bibfield  {author} {\bibinfo {author} {\bibfnamefont {N.}~\bibnamefont
  {Korabel}}\ and\ \bibinfo {author} {\bibfnamefont {E.}~\bibnamefont
  {Barkai}},\ }\bibfield  {title} {\enquote {\bibinfo {title} {Paradoxes of
  subdiffusive infiltration in disordered systems},}\ }\href@noop {} {\bibfield
   {journal} {\bibinfo  {journal} {Phys. Rev. Lett.}\ }\textbf {\bibinfo
  {volume} {104}},\ \bibinfo {pages} {170603} (\bibinfo {year}
  {2010})}\BibitemShut {NoStop}%
\bibitem [{\citenamefont {Straka}(2018)}]{straka2018variable}%
  \BibitemOpen
  \bibfield  {author} {\bibinfo {author} {\bibfnamefont {P.}~\bibnamefont
  {Straka}},\ }\bibfield  {title} {\enquote {\bibinfo {title} {Variable order
  fractional {F}okker--{P}lanck equations derived from continuous time random
  walks},}\ }\href@noop {} {\bibfield  {journal} {\bibinfo  {journal} {Physica
  A}\ }\textbf {\bibinfo {volume} {503}},\ \bibinfo {pages} {451} (\bibinfo
  {year} {2018})}\BibitemShut {NoStop}%
\bibitem [{\citenamefont {Fedotov}\ and\ \citenamefont
  {Han}(2019)}]{fedotov2019asymptotic}%
  \BibitemOpen
  \bibfield  {author} {\bibinfo {author} {\bibfnamefont {S.}~\bibnamefont
  {Fedotov}}\ and\ \bibinfo {author} {\bibfnamefont {D.}~\bibnamefont {Han}},\
  }\bibfield  {title} {\enquote {\bibinfo {title} {Asymptotic behavior of the
  solution of the space dependent variable order fractional diffusion equation:
  ultra-slow anomalous aggregation},}\ }\href@noop {} {\bibfield  {journal}
  {\bibinfo  {journal} {Phys. Rev. Lett.}\ }\textbf {\bibinfo {volume} {123}},\
  \bibinfo {pages} {050602} (\bibinfo {year} {2019})}\BibitemShut {NoStop}%
\bibitem [{\citenamefont {Fedotov}\ and\ \citenamefont
  {Falconer}(2012)}]{fedotov2012subdiffusive}%
  \BibitemOpen
  \bibfield  {author} {\bibinfo {author} {\bibfnamefont {S.}~\bibnamefont
  {Fedotov}}\ and\ \bibinfo {author} {\bibfnamefont {S.}~\bibnamefont
  {Falconer}},\ }\bibfield  {title} {\enquote {\bibinfo {title} {Subdiffusive
  master equation with space-dependent anomalous exponent and structural
  instability},}\ }\href@noop {} {\bibfield  {journal} {\bibinfo  {journal}
  {Phys. Rev. E}\ }\textbf {\bibinfo {volume} {85(3)}},\ \bibinfo {pages}
  {031132} (\bibinfo {year} {2012})}\BibitemShut {NoStop}%
\bibitem [{\citenamefont {Shkilev}(2013)}]{shkilev2013boundary}%
  \BibitemOpen
  \bibfield  {author} {\bibinfo {author} {\bibfnamefont {V.P.}\ \bibnamefont
  {Shkilev}},\ }\bibfield  {title} {\enquote {\bibinfo {title} {Boundary
  conditions for the subdiffusion equation},}\ }\href@noop {} {\bibfield
  {journal} {\bibinfo  {journal} {J. Exp. Theor. Phys.}\ }\textbf {\bibinfo
  {volume} {116}},\ \bibinfo {pages} {703} (\bibinfo {year}
  {2013})}\BibitemShut {NoStop}%
\bibitem [{\citenamefont {Koszto{\l}owicz}(2018)}]{kosztolowicz2018model}%
  \BibitemOpen
  \bibfield  {author} {\bibinfo {author} {\bibfnamefont {T.}~\bibnamefont
  {Koszto{\l}owicz}},\ }\bibfield  {title} {\enquote {\bibinfo {title} {Model
  of subdiffusion--absorption process in a membrane system consisting of two
  different media},}\ }\href@noop {} {\bibfield  {journal} {\bibinfo  {journal}
  {Acta Phys. Pol. B}\ }\textbf {\bibinfo {volume} {49}},\ \bibinfo {pages}
  {943} (\bibinfo {year} {2018})}\BibitemShut {NoStop}%
\bibitem [{\citenamefont
  {Koszto{\l}owicz}(2008)}]{kosztolowicz2008subdiffusion}%
  \BibitemOpen
  \bibfield  {author} {\bibinfo {author} {\bibfnamefont {T.}~\bibnamefont
  {Koszto{\l}owicz}},\ }\bibfield  {title} {\enquote {\bibinfo {title}
  {Subdiffusion in a system with a thick membrane},}\ }\href@noop {} {\bibfield
   {journal} {\bibinfo  {journal} {J. Membr. Sci.}\ }\textbf {\bibinfo {volume}
  {320}},\ \bibinfo {pages} {492} (\bibinfo {year} {2008})}\BibitemShut
  {NoStop}%
\bibitem [{\citenamefont {Koszto{\l}owicz}(2015)}]{kosztolowicz2015random}%
  \BibitemOpen
  \bibfield  {author} {\bibinfo {author} {\bibfnamefont {T.}~\bibnamefont
  {Koszto{\l}owicz}},\ }\bibfield  {title} {\enquote {\bibinfo {title} {Random
  walk model of subdiffusion in a system with a thin membrane},}\ }\href@noop
  {} {\bibfield  {journal} {\bibinfo  {journal} {Phys. Rev. E}\ }\textbf
  {\bibinfo {volume} {91}},\ \bibinfo {pages} {022102} (\bibinfo {year}
  {2015})}\BibitemShut {NoStop}%
\bibitem [{\citenamefont {Korabel}\ and\ \citenamefont
  {Barkai}(2011)}]{korabel2011boundary}%
  \BibitemOpen
  \bibfield  {author} {\bibinfo {author} {\bibfnamefont {N.}~\bibnamefont
  {Korabel}}\ and\ \bibinfo {author} {\bibfnamefont {E.}~\bibnamefont
  {Barkai}},\ }\bibfield  {title} {\enquote {\bibinfo {title} {Boundary
  conditions of normal and anomalous diffusion from thermal equilibrium},}\
  }\href@noop {} {\bibfield  {journal} {\bibinfo  {journal} {Phys. Rev. E}\
  }\textbf {\bibinfo {volume} {83}},\ \bibinfo {pages} {051113} (\bibinfo
  {year} {2011})}\BibitemShut {NoStop}%
\bibitem [{\citenamefont {Marseguerra}\ and\ \citenamefont
  {Zoia}(2006)}]{marseguerra2006normal}%
  \BibitemOpen
  \bibfield  {author} {\bibinfo {author} {\bibfnamefont {M.}~\bibnamefont
  {Marseguerra}}\ and\ \bibinfo {author} {\bibfnamefont {A.}~\bibnamefont
  {Zoia}},\ }\bibfield  {title} {\enquote {\bibinfo {title} {Normal and
  anomalous transport across an interface: Monte carlo and analytical
  approach},}\ }\href@noop {} {\bibfield  {journal} {\bibinfo  {journal} {Ann.
  Nucl. Energy}\ }\textbf {\bibinfo {volume} {33}},\ \bibinfo {pages} {1396}
  (\bibinfo {year} {2006})}\BibitemShut {NoStop}%
\bibitem [{\citenamefont {Klafter}\ and\ \citenamefont
  {Sokolov}(2011)}]{steps}%
  \BibitemOpen
  \bibfield  {author} {\bibinfo {author} {\bibfnamefont {J.}~\bibnamefont
  {Klafter}}\ and\ \bibinfo {author} {\bibfnamefont {I.M.}\ \bibnamefont
  {Sokolov}},\ }\href@noop {} {\emph {\bibinfo {title} {First Steps in Random
  Walks}}}\ (\bibinfo  {publisher} {Oxford Univ. Press, Oxford},\ \bibinfo
  {year} {2011})\BibitemShut {NoStop}%
\bibitem [{\citenamefont {Barkai}\ and\ \citenamefont
  {Sokolov}(2007)}]{BarSok}%
  \BibitemOpen
  \bibfield  {author} {\bibinfo {author} {\bibfnamefont {E.}~\bibnamefont
  {Barkai}}\ and\ \bibinfo {author} {\bibfnamefont {I.M.}\ \bibnamefont
  {Sokolov}},\ }\bibfield  {title} {\enquote {\bibinfo {title} {On {H}ilfer's
  objection to the fractional time diffusion equation},}\ }\href@noop {}
  {\bibfield  {journal} {\bibinfo  {journal} {Physica A}\ }\textbf {\bibinfo
  {volume} {373}},\ \bibinfo {pages} {231} (\bibinfo {year}
  {2007})}\BibitemShut {NoStop}%
\bibitem [{\citenamefont {Barenblatt}\ and\ \citenamefont
  {Zel'dovich}(1972)}]{BarZel}%
  \BibitemOpen
  \bibfield  {author} {\bibinfo {author} {\bibfnamefont {G.I.}\ \bibnamefont
  {Barenblatt}}\ and\ \bibinfo {author} {\bibfnamefont {Ya.B.}\ \bibnamefont
  {Zel'dovich}},\ }\bibfield  {title} {\enquote {\bibinfo {title} {Self-similar
  solutions as intermediate asymptotics},}\ }\href@noop {} {\bibfield
  {journal} {\bibinfo  {journal} {Annu. Rev. Fluid Mech.}\ }\textbf {\bibinfo
  {volume} {4}},\ \bibinfo {pages} {285} (\bibinfo {year} {1972})}\BibitemShut
  {NoStop}%
\bibitem [{\citenamefont {Barenblatt}(1996)}]{Barenblatt}%
  \BibitemOpen
  \bibfield  {author} {\bibinfo {author} {\bibfnamefont {G.I.}\ \bibnamefont
  {Barenblatt}},\ }\href@noop {} {\emph {\bibinfo {title} {Scaling,
  self-similarity, and intermediate asymptotics: dimensional analysis and
  intermediate asymptotics}}}\ (\bibinfo  {publisher} {Cambridge University
  Press, Cambridge, England},\ \bibinfo {year} {1996})\BibitemShut {NoStop}%
\bibitem [{\citenamefont {West}\ \emph {et~al.}(2012)\citenamefont {West},
  \citenamefont {Bologna},\ and\ \citenamefont {Grigolini}}]{PFO}%
  \BibitemOpen
  \bibfield  {author} {\bibinfo {author} {\bibfnamefont {B.}~\bibnamefont
  {West}}, \bibinfo {author} {\bibfnamefont {M.}~\bibnamefont {Bologna}}, \
  and\ \bibinfo {author} {\bibfnamefont {P.}~\bibnamefont {Grigolini}},\
  }\href@noop {} {\emph {\bibinfo {title} {Physics of fractal operators}}}\
  (\bibinfo  {publisher} {Springer, New York},\ \bibinfo {year}
  {2012})\BibitemShut {NoStop}%
\bibitem [{\citenamefont {Gorenflo}\ \emph {et~al.}(2007)\citenamefont
  {Gorenflo}, \citenamefont {Mainardi},\ and\ \citenamefont
  {Vivoli}}]{Gorenflo}%
  \BibitemOpen
  \bibfield  {author} {\bibinfo {author} {\bibfnamefont {R.}~\bibnamefont
  {Gorenflo}}, \bibinfo {author} {\bibfnamefont {F.}~\bibnamefont {Mainardi}},
  \ and\ \bibinfo {author} {\bibfnamefont {A.}~\bibnamefont {Vivoli}},\
  }\bibfield  {title} {\enquote {\bibinfo {title} {Continuous-time random walk
  and parametric subordination in fractional diffusion},}\ }\href@noop {}
  {\bibfield  {journal} {\bibinfo  {journal} {Chaos, Solitons \& Fractals}\
  }\textbf {\bibinfo {volume} {34}},\ \bibinfo {pages} {87} (\bibinfo {year}
  {2007})}\BibitemShut {NoStop}%
\bibitem [{\citenamefont {Sokolov}(2010)}]{ItoStrat}%
  \BibitemOpen
  \bibfield  {author} {\bibinfo {author} {\bibfnamefont {I.M.}\ \bibnamefont
  {Sokolov}},\ }\bibfield  {title} {\enquote {\bibinfo {title} {Ito,
  {S}tratonovich, {H}{\"a}nggi and all the rest: The thermodynamics of
  interpretation},}\ }\href@noop {} {\bibfield  {journal} {\bibinfo  {journal}
  {Chem. Phys.}\ }\textbf {\bibinfo {volume} {375}},\ \bibinfo {pages} {359}
  (\bibinfo {year} {2010})}\BibitemShut {NoStop}%
\bibitem [{\citenamefont {Kuhlmann}(2013)}]{gaver}%
  \BibitemOpen
  \bibfield  {author} {\bibinfo {author} {\bibfnamefont {K.~L.}\ \bibnamefont
  {Kuhlmann}},\ }\bibfield  {title} {\enquote {\bibinfo {title} {Review on
  inverse {L}aplace transform algorithms for {L}aplace-space numerical
  approaches},}\ }\href@noop {} {\bibfield  {journal} {\bibinfo  {journal}
  {Numer. Algorithms}\ }\textbf {\bibinfo {volume} {63}},\ \bibinfo {pages}
  {339} (\bibinfo {year} {2013})}\BibitemShut {NoStop}%
\bibitem [{\citenamefont {Prudnikov}\ \emph {et~al.}(1990)\citenamefont
  {Prudnikov}, \citenamefont {Brychkov},\ and\ \citenamefont
  {Marichev}}]{prudnikovyu}%
  \BibitemOpen
  \bibfield  {author} {\bibinfo {author} {\bibfnamefont {A.P.}\ \bibnamefont
  {Prudnikov}}, \bibinfo {author} {\bibfnamefont {Yu.A.}\ \bibnamefont
  {Brychkov}}, \ and\ \bibinfo {author} {\bibfnamefont {O.I.}\ \bibnamefont
  {Marichev}},\ }\href@noop {} {\emph {\bibinfo {title} {Integrals and Series,
  Vol. 3: More Special Functions}}}\ (\bibinfo  {publisher} {Gordon and Breach,
  New York},\ \bibinfo {year} {1990})\BibitemShut {NoStop}%
\end{thebibliography}
\end{document}